\documentclass[journal=langd5,manuscript=letter,layout=twocolumn]{achemso}
\usepackage{graphicx}% Include figure files
\usepackage{amsmath}
\usepackage{amssymb}
\newcommand{\un}[1]{\ensuremath{\unskip\,\mathrm{#1}}}
\newcommand{\form}{\mbox{C$_{12}$EO$_{5} \,$}}

\title{Lyotropic lamellar phase doped with a nematic phase of magnetic nanorods}

\author{Doru Constantin}
\email{constantin@lps.u-psud.fr}
\affiliation{Laboratoire de Physique des Solides, Universit\'{e}
Paris-Sud, CNRS, UMR8502, 91405 Orsay, France.}

\author{Patrick Davidson}
\affiliation{Laboratoire de Physique des Solides, Universit\'{e}
Paris-Sud, CNRS, UMR8502, 91405 Orsay, France.}

\author{Corinne Chan\'{e}ac}
\affiliation{Laboratoire de Chimie de la Mati\`{e}re Condens\'{e}e,
Universit\'{e} Paris VI, CNRS, UMR7574, 75252 Paris, France.}

\begin{document}

\begin{abstract}
We report the elaboration of a hybrid mesophase combining the
lamellar order of a lyotropic system of nonionic surfactant and the
nematic order of a concentrated solution of inorganic nanorods
confined between the surfactant layers. Highly aligned samples of
this mesophase can be obtained by thermal annealing, and the
orientation of the nanorods is readily controlled with a magnetic
field. High-resolution synchrotron x-ray scattering and polarised
optical microscopy show that, compared to their isolated
counterparts, both the nematic and lamellar orders are altered,
demonstrating their interplay.
\end{abstract}

%\pacs{61.30.St, 61.05.cf, 82.70.Dd}

%     61.30.St    Lyotropic phases
%     61.05.cf    X-ray scattering (including small-angle scattering)
%     82.70.Dd    Colloids

\bigskip

\bigskip

%\centerline{\includegraphics[width=3.25in]{TOC.eps}}

\newpage

%\maketitle

%\section*{Introduction}

Hybrid soft-matter systems, combining at the nanometric scale two
components with different types of order, have recently been the
focus of a fast-increasing body of research. Indeed, such systems
raise interesting fundamental questions about the interaction of the
different kinds of order involved and also offer new
perspectives of applications, for example in drug
delivery \cite{Ewert04}.

In particular, lamellar phases doped with anisotropic inclusions received sustained attention since their experimental realization in DNA/lipid complexes \cite{Raedler97,Koltover98}. Since then, various other organic dopants have been used, such as viruses \cite{Yang04} and peptides \cite{Koltover04}. These experimental achievements also prompted extensive theoretical efforts \cite{Salditt97,Golubovic98,OHern98}, concentrating on the interplay of the two types of order (the lamellar one of the matrix and that of the inclusions within it), and on the characteristics of a possible 2D-smectic phase formed by the inclusions.

These systems are generally obtained by electrostatic complexation
through a precipitation process; as such, their texture is very hard
to control and they usually occur in the form of multilayered
globules with random orientation. It is therefore difficult to
discriminate between their molecular organization in the plane of
the layers and that along the director of the phase. Furthermore,
they are quite concentrated (lamellar spacings of a few~nm), which
severely restricts the size of the inclusions.

In this Letter, we present a new hybrid system,
consisting of a dilute lamellar phase (formed by a nonionic
surfactant) doped with a nematic phase of inorganic goethite
($\alpha-\un{FeOOH}$) nanorods that differ from the organic dopants
used so far by their magnetic properties. The resulting composite
(nematic/lamellar) phase is very fluid, and hence easily aligned,
which allowed us to study its structure in detail. A notable
advantage is that the orientation of the nanorods can be controlled
by an externally applied magnetic field of moderate strength.

The degree of order of each component (quantified by the width of
the Bragg peak for the lamellar host phase and by the nematic order
parameter for the confined nanorods) changes due to the presence of
the other component, confirming their intimate interaction.

Furthermore, this system shows promise as template for the
production of hybrid materials, e.g. for magnetic storage
applications \cite{Sun00}, shielding \cite{Gass06},
metamaterials \cite{Shelby01} etc.

%\section*{Experimental}

The matrix is the C$_{12}$EO$_{5}$/hexanol/H$_{2}$O$\,$ system,
where C$_{12}$EO$_{5}$ stands for the nonionic surfactant
penta(ethylene glycol) monododecyl ether. Its lamellar phase can be
diluted down to spacings $d$ in the micron range, while the bilayer
thickness $\delta \approx 2.9 \un{nm}$
\cite{Freyssingeas96,Freyssingeas97}. We used a hexanol/\form ratio
of 0.33 by weight, corresponding to a molar ratio of 1.3 (hexanol
molecules for each surfactant molecule). The main role of hexanol is
to bring the lamellar phase domain down to room temperature. The
surfactant was acquired from Nikko and the 1--hexanol from Fluka; they
were used without further purification. For all the samples discussed in this paper, the fraction of membrane $\phi_m = (V_{\scriptsize{\form}} + V_{\mathrm{hexanol}})/V_{\mathrm{total}} = 6.26$~vol~\%.

Goethite ($\alpha-\un{FeOOH}$) is widely used as a pigment of ochre
color \cite{Cornell96}. In bulk, its density is $\rho_g = 4.37 \un{g/cm^3}$. The nanorods were synthesized according to
well-established protocols \cite{Atkinson67,Jolivet04}. Stable
aqueous suspensions of non-aggregated goethite nanorods are obtained
by repeated centrifugation and dispersion in water up to $\un{pH} =
3$, where their surface is hydroxylated, with a surface charge of
$0.2 \un{C \, m^{-2}}$ (the isoelectric point corresponds to
$\un{pH} = 9$). Although bulk goethite is antiferromagnetic, the
nanorods bear a permanent magnetic dipole $\mu \sim 1200 \, \mu
_{B}$ along their long axis, probably due to uncompensated surface
spins (with $\mu _{B} = 9.274 \, 10^{-24}\, \un{J/T}$ the Bohr
magneton). Therefore, in suspension, the nanorods are easily aligned
parallel to a small magnetic field. Furthermore, the easy
magnetisation axis is perpendicular to this direction so that, at
high applied fields, the induced magnetic moment overtakes the
permanent one and the orientation of the rods switches to
perpendicular to the field at a critical value $B \sim 350 \un{mT}$
\cite{Lemaire02}. When the rods are confined within the lamellar phase, the reorientation also occurs, at the same field value, and the texture of the lamellar phase follows the orientation of the rods \cite{Constantin08}.

TEM observations were made on deposits of one drop of dilute
nanoparticle suspension on a copper grid covered with a carbon
membrane. The morphology is typical for a goethite crystal,
elongated along the [001] direction and terminated by \{210\} faces
\cite{Cornell96} with an aspect ratio of 8.4. The particle size
distribution was determined over a population of 200 particles. The
average length is $\bar{L}= 315 \un{nm}$ and the standard deviation
$\sigma_L = 88 \un{nm}$. For the transverse dimension, $\bar{D} = 42
\un{nm}$ and $\sigma_D = 12 \un{nm}$. The polydispersities are
relatively low, $\sigma_L /\bar{L} = 0.28$ and $\sigma_D /\bar{D} =
0.3$. More precisely, the particles are lath-shaped and from powder
X-ray diffraction line-broadening we infer that they have a mean
width of 38~nm and a mean thickness of 18~nm, in good agreement with
the TEM results.

Small angle x-ray scattering (SAXS) experiments were performed at
the ID02 station of the European Synchrotron Radiation Facility
synchrotron in Grenoble, France. The incident beam had a wavelength
$\lambda = 0.0995\,\un{nm}$, and the sample--detector distance was
5~m. The scattered x-rays were detected with a specially developed
CCD camera. A detailed description of the experimental setup can be
found in reference 19. The $q$ range over which the
data could be reliably collected was $0.02 < q < 0.6 \un{nm}^{-1}$.
The samples were held in flat glass capillaries, 50~$\mu$m thick and
1~mm wide (Vitrocom, NJ, USA). The flat faces of the capillaries
were set perpendicular to the x-ray beam. The magnetic field was
applied using a motorized variable-gap setup available at the
beamline.

To determine the order parameter of the nematic phase, azimuthal
sections $I(\theta)$ through the scattered signal at the radial
position $q_{max} = 2 \pi /(80 \un{nm})$ of the nematic peak were
fitted with a profile derived from the Maier-Saupe theory, as
discussed in detail in references 20--22.

\begin{figure*}[htbp]
\includegraphics[width=0.7\textwidth,angle=0]{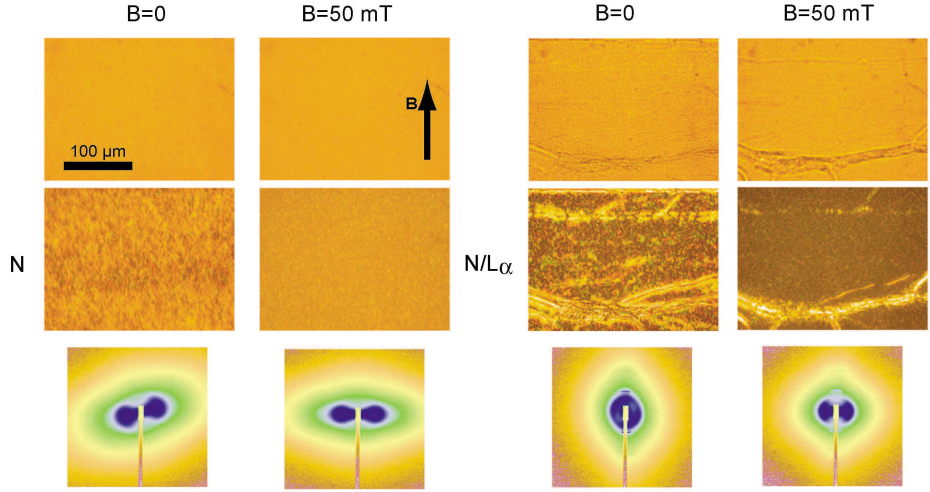}
\caption{Optical microscopy textures (top) and SAXS signal (bottom)
of the nematic phase of goethite, at a concentration
$\phi_g=8$~vol~\%, in water (left) and contained within the lamellar
$L_{\alpha}$ phase (right). The lower microscopy images are taken
between crossed polarizers, parallel to the image sides. In both
cases, the nematic phase is very well aligned along the magnetic
field, even at a relatively weak value of 50~mT. In the panel on the
right, the lamellar phase is almost completely aligned in
homeotropic anchoring (bilayers parallel to the flat faces of the
capillary), with the exception of a few oily streak defects, visible
in the microscopy images and which give rise to the very weak and
sharp peaks along the vertical axis in the SAXS
images.}\label{fig:fig1}
\end{figure*}

Microscopy observations were done using an Olympus BX51 microscope
(at $5\times$--$40\times$ magnification) using linearly polarized
light and, when specified, an analyzer perpendicular to the incident
polarization. For birefringence measurements we used a Berek
compensator (U-CBE from Olympus) and a green band-pass filter
(480--580~nm.) The magnetic field was applied using a home-made
setup based on permanent magnets with a variable gap. One can thus
reach field intensities of up to 0.9~T.

%\section*{Results and discussion}

\ref{fig:fig1} presents a comparison between an aqueous
nematic suspension $N$ of goethite nanorods (left) and the hybrid
nematic/lamellar ($N/L_{\alpha}$) mesophase (right), with and
without an applied magnetic field. For the $N/L_{\alpha}$ system, in
the absence of a magnetic field, the texture of the phase as
observed between crossed polarizers (middle row) exhibits both the
oily-streak defects due to the smectic symmetry of the lamellar
component and textures typical for a nematic phase (in-between the oily streaks). Applying a modest (50~mT) magnetic
field aligns the nematic component, so that only the lamellar
defects remain visible.

\begin{figure*}[htbp]
\includegraphics[width=0.8\textwidth,angle=0]{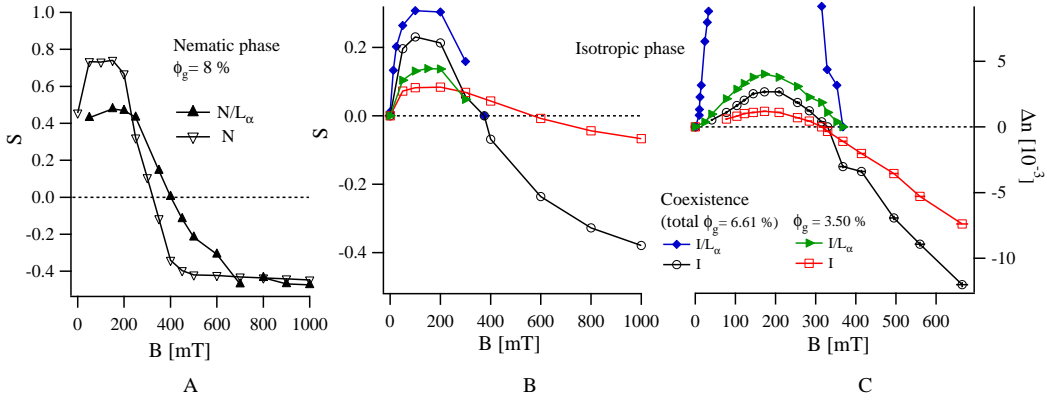}
\caption{A -- Order parameter $S$ (calculated from the SAXS
patterns) of the nematic phase of goethite nanorods, in the lamellar
phase ($\blacktriangle$) and in water ($\triangledown$) as a
function of the applied magnetic field. At field values $B\gtrsim
350 \un{mT}$ the nanorods align perpendicular to the field,
resulting in negative values of the order parameter. B -- Induced
order parameter and C -- induced birefringence of the isotropic
phase of goethite nanorods as a function of the applied magnetic
field, for two concentrations $\phi _g$. The symbols are the same in
subfigures B and C. The optical birefringence measurements cannot be performed in the lamellar phase above the reorientation field value due to the change in texture (homeotropic to planar).}\label{fig:fig4}
\end{figure*}

Using SAXS, we studied the orientation of the goethite nanorods
confined in the $N/L_{\alpha}$ phase. The SAXS patterns (\ref{fig:fig1}, bottom row) show that the confined nanorods are
easily aligned along the magnetic field direction, without
disrupting the texture of the lamellar phase. The particles remain
aligned when the magnetic field is switched off, showing that the
nematic uniform alignment of the confined nanorods is stable.
Moreover, we measured by SAXS the order parameter $S$ in the $N$ and
$N/L_{\alpha}$ phases at the same particle concentration $\phi _g =
8$~vol~\%, on samples aligned using a moderate magnetic field
(50--200~mT). While the aqueous nematic phase $N$ has $S \simeq
0.75$, the hybrid system exhibits lower nematic order, with $S
\simeq 0.45$ (\ref{fig:fig4}A). This decrease could be due to
the weakness of orientational correlations between the particles
confined between different surfactant bilayers. Above the reorientation threshold, the order parameter of the rods varies continuously and reaches saturation at about 700~mT (\ref{fig:fig4}A). No significant effect was observed above this value. Also, the magnetic field has no detectable effect on the undoped lamellar phase, even at the highest field values we could reach.

Another noteworthy characteristic of the hybrid system is the
enhanced susceptibility of the isotropic confined particle phase
($I/L_{\alpha}$), quantified by the induced order parameter $S(B)$
(\ref{fig:fig4}B) and birefringence $\Delta n (B)$ (\ref{fig:fig4}C) under an applied magnetic field $B$. The experiments were performed for two different goethite
concentrations, $\phi _g = 3.5$~vol~\% and at coexistence with the
$N/L_{\alpha}$ phase, at an overall concentration  $\phi _g =
6.61$~vol~\%. Due to the presence of the lamellar phase, we were not able to determine precisely the goethite content in the two coexisting phases, but it appears to be roughly similar to that in aqueous solutions (approx. 4.5~:~7.5~vol~\%).

\begin{figure*}[htbp]
\includegraphics[width=0.8\textwidth,angle=0]{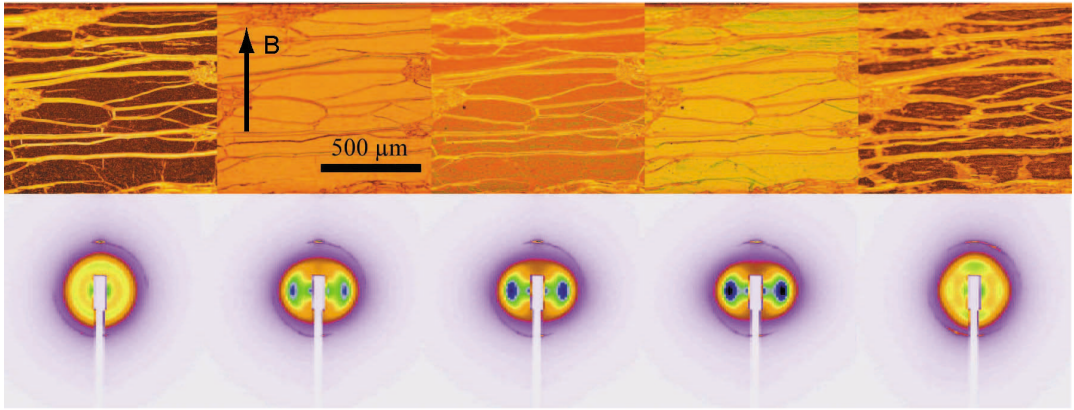}
\caption{Optical microscopy textures (top) and SAXS signal (bottom)
of the lamellar $L_{\alpha}$ phase doped with the goethite isotropic
phase ($I/L_{\alpha}$) at coexistence with the nematic-doped phase
$N/L_{\alpha}$ for different values of the applied field. The
microscopy images are taken between crossed polarizers, parallel to
the image sides. From left to right, the field values are: $B=0$,
27, 52.5, 104 and 370~mT.}\label{fig:fig2}
\end{figure*}

Both parameters are clearly higher in the hybrid
system. This feature is extremely strong for the field-induced
birefringence (diamonds in \ref{fig:fig4}C) of a
confined isotropic phase of volume fraction $\phi _g = 6.61$~vol~\%,
within the biphasic domain of aqueous goethite suspensions. At this concentration, there is coexistence between the isotropic- and nematic-doped lamellar phases ($I/L_{\alpha}$ and $N/L_{\alpha}$), which were identified optically, within the same capillary, by their distinct textures. The gap
between 33 and 350~mT corresponds to birefringence values that
cannot be reliably determined using our setup. The corresponding
microscopy and SAXS images are shown in \ref{fig:fig2}, where
the strong birefringence is revealed by the color shift and the
progressively increasing order parameter of the nanorods by the
horizontal lobes in the scattering pattern.

Both the lower order parameter in the nematic phase and the higher
susceptibility in the isotropic phase are compatible with a
second-order phase transition (predicted in the literature for a 2D nematic phase \cite{vanderSchoot97,Wink07}), as opposed to the first-order
transition in the aqueous system \cite{Lemaire02}.

\begin{figure}[htbp]
\includegraphics[width=0.4\textwidth,angle=0]{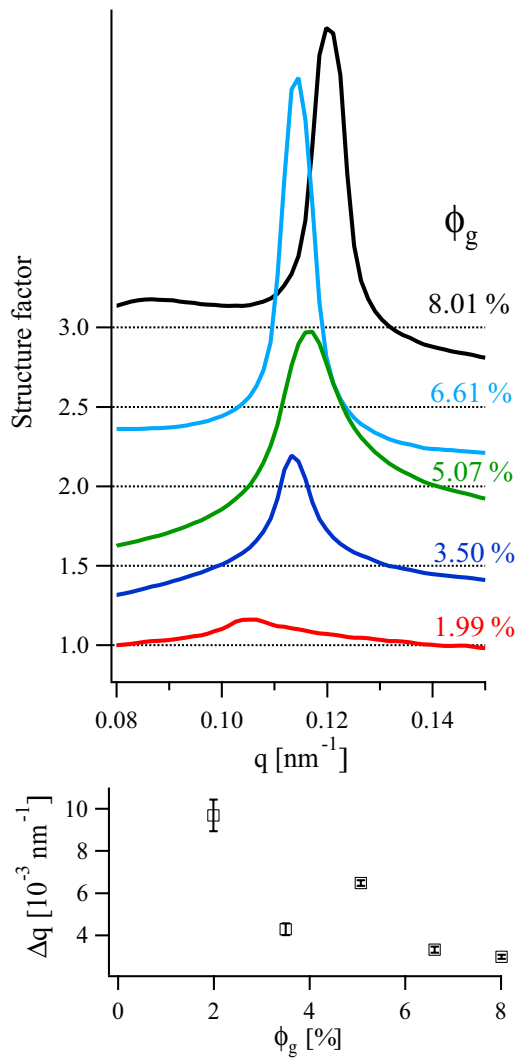}
\caption{A -- SAXS image (taken without magnetic field) of a
lamellar phase containing $\phi _g = 6.61$~vol~\% goethite, aligned
in planar anchoring (the layer normal is along $z$.) Several Bragg
peaks are visible, labelled by their index. B -- Structure factors
of the lamellar phase doped with increasing amounts of goethite,
indicated alongside the curves. The curves are shifted vertically in
steps of 0.5. C -- Width of the Bragg peaks in B as a function of
the goethite concentration.}\label{fig:fig3}
\end{figure}

Finally, the presence of the inclusions affects the structure of the
host lamellar phase, which becomes stiffer, as seen by the decreasing
width of the Bragg peak with increasing goethite concentration. This
effect is displayed in \ref{fig:fig3}. We can tentatively
attribute it to a strong interaction between the nanorods and the
surfactant bilayers, leading to the formation of hydrogen bonds
\cite{Frost05}. The nanorods are thus adsorbed onto the bilayers and
increase their stiffness.

We emphasize that the two components (the surfactant layers and the nanorods) are intimately mixed. While the lamellar order is of course imposed by the surfactant bilayers, it also applies to the nanorods confined between these bilayers. As a result, there is only one repeat distance, giving rise to the single set of Bragg peaks, discussed in \ref{fig:fig3}.

It is noteworthy that the overall X-ray signal is mainly due to the
nanorods (at this dilution, the contribution of the surfactant
bilayers is negligible). Indeed, the structure factors are obtained
dividing the measured intensity by the form factor of the nanorods \cite{Note-1}.
%\bibnote{The form factor was measured in a dilute isotropic
%solution, so it corresponds to randomly oriented nanorods. However,
%for highly anisotropic particles and as long as the scattering
%vector is not too small, this is dominated by the transverse form
%factor of the particles}. This detail is relevant insofar it
%emphasizes that the host phase transmits its lamellar order to the
%doping particles.
No peak can be detected for goethite concentrations $\phi _g < 2$~vol~\%.

%\section*{Conclusion}

The two components of the hybrid mesophase interact in non-trivial
ways, as demonstrated by the enhanced magnetic field susceptibility
of the nanorods (\ref{fig:fig4}, B and C) and by the
increased stiffness of the lamellar phase (\ref{fig:fig3}).
For the most concentrated system, the nanorods exhibit both nematic
order (also encountered in aqueous solutions) and a lamellar order
imposed by the confining surfactant bilayers. In this respect, the
phase is similar to those encountered in complexes formed by DNA
with cationic lipids, with the important distinction that we use a
dilute phase of nonionic surfactant, which is easily aligned by
thermal treatment and that the orientation of the nanorods couples
to an external magnetic field. As such, this hybrid phase could
provide an ideal testing ground for the hypothesized ``sliding
phases'' \cite{OHern99}, stacks of weakly-coupled layers with a
certain degree of two-dimensional (in-plane) order within the
layers.

From a practical point of view, the combination of these two types
of order makes the system a promising candidate for the formulation
of composite materials with controlled periodicity and anisotropy,
ordered over macroscopic distances. Moreover, the magnetic
properties of the goethite nanorods are particularly interesting in
this respect.

\newpage

\begin{acknowledgement}
The ESRF is acknowledged for the provision of beamtime (experiment
SC-2393, ID02 beamline). The authors thank P. Boesecke, M.
Imp\'{e}ror, A. Poulos and B. Pansu for assistance with the
synchrotron SAXS experiments and S. Rouzi\`{e}re for technical help.
\end{acknowledgement}

%\bibliography{nem_lam,goeth_ref}

\begin{thebibliography}{10}

\bibitem{Ewert04}
Ewert,~K.;\ \ Slack,~N.~L.;\ \ Ahmad,~A.;\ \ Evans,~H.~M.;\ \ Lin,~A.~J.;\ \
  Samuel,~C.~E.;\ \ Safinya,~C.~R. \textit{Current Medicinal Chemistry}
  \textbf{2004,} \textsl{11,} 133-149.

\bibitem{Raedler97}
R\"{a}dler,~J.~O.;\ \ Koltover,~I.;\ \ Salditt,~T.;\ \ Safinya,~C.~R.
  \textit{Science} \textbf{1997,} \textsl{275,} 810-814.

\bibitem{Koltover98}
Koltover,~I.;\ \ Salditt,~T.;\ \ R\"{a}dler,~J.~O.;\ \ Safinya,~C.~R.
  \textit{Science} \textbf{1998,} \textsl{281,} 78-81.

\bibitem{Yang04}
Yang,~L.;\ \ Liang,~H.;\ \ Angelini,~T.~E.;\ \ Butler,~J.;\ \ Coridan,~R.;\ \
  Tang,~J.~X.;\ \ Wong,~G. C.~L. \textit{Nature Materials} \textbf{2004,}
  \textsl{3,} 615-619.

\bibitem{Koltover04}
Koltover,~I.;\ \ Sahu,~S.;\ \ Davis,~N. \textit{Angew. Chem.} \textbf{2004,}
  \textsl{116,} 4126-4129.

\bibitem{Salditt97}
Salditt,~T.;\ \ Koltover,~I.;\ \ R\"{a}dler,~J.~O.;\ \ Safinya,~C.~R.
  \textit{Phys. Rev. Lett.} \textbf{1997,} \textsl{79,} 2582-2585.

\bibitem{Golubovic98}
Golubovi\v{c},~L.;\ \ Golubovi\v{c},~M. \textit{Phys. Rev. Lett.}
  \textbf{1998,} \textsl{80,} 4341-4344.

\bibitem{OHern98}
{O'Hern},~C.~S.;\ \ Lubensky,~T.~C. \textit{Phys. Rev. Lett.} \textbf{1998,}
  \textsl{80,} 4345-4348.

\bibitem{Sun00}
Sun,~S.;\ \ Murray,~C.~B.;\ \ Weller,~D.;\ \ Folks,~L.;\ \ Moser,~A.
  \textit{Science} \textbf{2000,} \textsl{287,} 1989-1992.

\bibitem{Gass06}
Gass,~J.;\ \ Poddar,~P.;\ \ Almand,~J.;\ \ Srinath,~S.;\ \ Srikanth,~H.
  \textit{Adv. Funct. Mater.} \textbf{2006,} \textsl{16,} 71-75.

\bibitem{Shelby01}
Shelby,~R.~A.;\ \ Smith,~D.~R.;\ \ Schultz,~S. \textit{Science} \textbf{2001,}
  \textsl{292,} 77-79.

\bibitem{Freyssingeas96}
Freyssingeas,~E.;\ \ Nallet,~F.;\ \ Roux,~D. \textit{Langmuir} \textbf{1996,}
  \textsl{12,} 6028-6035.

\bibitem{Freyssingeas97}
Freyssingeas,~E.;\ \ Roux,~D.;\ \ Nallet,~F. \textit{J. Phys. {II} (France)}
  \textbf{1997,} \textsl{7,} 913-929.

\bibitem{Cornell96}
Cornell,~R.~M.;\ \ Schwertmann,~U. \textit{The iron oxides, structure,
  properties, reactions, occurrence and uses;} VCH: Weinheim, 1996.

\bibitem{Atkinson67}
Atkinson,~R.~J.;\ \ Posner,~A.~M.;\ \ Quirk,~J.~P. \textit{J. Phys. Chem.}
  \textbf{1967,} \textsl{71,} 550-558.

\bibitem{Jolivet04}
Jolivet,~J.-P.;\ \ Chan\'{e}ac,~C.;\ \ Tronc,~E. \textit{Chem. Commun.}
  \textbf{2004,}  481-487.

\bibitem{Lemaire02}
Lemaire,~B.~J.;\ \ Davidson,~P.;\ \ Ferr\'{e},~J.;\ \ Jamet,~J.~P.;\ \
  Panine,~P.;\ \ Dozov,~I.;\ \ Jolivet,~J.~P. \textit{Phys. Rev. Lett.}
  \textbf{2002,} \textsl{88,} 125507.

\bibitem{Constantin08}
B\'{e}neut,~K.;\ \ Constantin,~D.;\ \ Davidson,~P.;\ \ Dessombz,~A.;\ \
  Chan\'{e}ac,~C. \textit{Langmuir} \textbf{2008,} \textsl{24,} 8205-8209.

\bibitem{Narayanan01}
Narayanan,~T.;\ \ Diat,~O.;\ \ Bosecke,~P. \textit{Nucl. Instrum. Methods Phys.
  Res. A} \textbf{2001,} \textsl{467,} 1005-1009.

\bibitem{Davidson95}
Davidson,~P.;\ \ Petermann,~D.;\ \ Levelut,~A.~M. \textit{J. Phys. II (France)}
  \textbf{1995,} \textsl{5,} 113-131.

\bibitem{Lemaire04b}
Lemaire,~B.~J.;\ \ Davidson,~P.;\ \ Ferr\'{e},~J.;\ \ Jamet,~J.~P.;\ \
  Petermann,~D.;\ \ Panine,~P.;\ \ Dozov,~I.;\ \ Jolivet,~J.~P. \textit{Eur.
  Phys. J. E} \textbf{2004,} \textsl{13,} 291-308.

\bibitem{Lemaire04c}
Lemaire,~B.~J.;\ \ Davidson,~P.;\ \ Petermann,~D.;\ \ Panine,~P.;\ \
  Dozov,~I.;\ \ Stoenescu,~D.;\ \ Jolivet,~J.~P. \textit{Eur. Phys. J. E}
  \textbf{2004,} \textsl{13,} 309-319.

\bibitem{vanderSchoot97}
van~der Schoot,~P. \textit{J. Chem. Phys.} \textbf{1997,} \textsl{106,}
  2355-2359.

\bibitem{Wink07}
Wink,~R. L.~C. \textit{Phys. Rev. Lett.} \textbf{2007,} \textsl{98,} 217801.

\bibitem{Frost05}
Frost,~R.;\ \ Zhu,~H.~Y.;\ \ Wu,~P.;\ \ Bostrom,~T. \textit{Materials Letters}
  \textbf{2005,} \textsl{59,} 2238-2241.

\bibitem{Note-1}
The form factor was measured in a dilute isotropic solution, so it corresponds
  to randomly oriented nanorods. However, for highly anisotropic particles and
  as long as the scattering vector is not too small, this is dominated by the
  transverse form factor of the particles.

\bibitem{OHern99}
{O'Hern},~C.~S.;\ \ Lubensky,~T.~C.;\ \ Toner,~J. \textit{Phys. Rev. Lett.}
  \textbf{1999,} \textsl{83,} 2745-2748.

\end{thebibliography}

%\newpage
%
%\bigskip
%\begin{center}
%\textbf{\LARGE For Table of Contents Use Only}
%\bigskip
%
%\includegraphics[width=3.25in]{TOC.eps}
%
%Lyotropic lamellar phase doped with a nematic phase of magnetic nanorods, by Doru Constantin, Patrick Davidson, and Corinne Chan\'{e}ac
%
%\end{center}

\end{document}